\documentclass[twocolumn,showpacs,preprintnumbers,amsmath,amssymb]{revtex4}

\usepackage{longtable}

\usepackage{graphicx}
\usepackage{amssymb}
\usepackage{epstopdf}
\usepackage{url}
\usepackage{mathpazo}

   \usepackage{rotating}
\newcommand{\bit}{\begin{itemize}} \newcommand{\eit}{\end{itemize}}

\newcommand{\ignore}[1]{}
\newcommand{\be}{\begin{equation}} \newcommand{\ee}{\end{equation}}
\newcommand{\ba}{\begin{eqnarray}} \newcommand{\ea}{\end{eqnarray}}
\newcommand{\nn}{\nonumber} \renewcommand{\bf}{\textbf}
\newcommand{\ra}{\rightarrow}

\newcommand{\p}{\partial}

\def\ket#1{\left| #1\right>}

\input{epsf}

\begin{document}

\title{Where and How to find $susy$:  \\ The auxiliary field interpretation of supersymmetry}
\author{John P. Ralston} 
\affiliation{Department of Physics \& Astronomy, University of Kansas, Lawrence KS 66045}

\begin{abstract}
 The {\it gauge hierarchy problem} found in perturbation theory is one of the main attractions for supersymmetry. Yet the quantum mechanical coupling of a low energy system to a high energy one invariably leads to {\it perturbative instability}, which is not a valid signal of dynamical inconsistency. We show by examples how perturbation theory with widely separated scales gives false results. We also identify the flaw in perturbative fine-tuning arguments. Non-perturbative features of random subsystems maintain and preserve the hierarchy in which they are embedded. After reviewing the likelihood the hierarchy problem is a perturbative fiction, we suggest a new interpretation of $susy$ as practical auxiliary fields. Their function is much like Feynman's gauge ghosts, developed in perturbation theory to repair illnesses of perturbation theory. $susy$ will be found useful when it is considered a tool of applied mathematics and data-fitting. We propose that $susy$ data fits should be customized to the particular experimental situations they are suited to improve, without dilution from the needless assumption that $susy$ must describe universal new physics. It is likely that $susy$ will soon be discovered a useful part of data analysis and diagnostics towards improving the understanding of the Standard Model, and possibly towards discovering what may constitute new physics after all.
 
\end{abstract}

%\pacs{}

\maketitle

\section*{Rationale for Supersymmetry}

\paragraph{Is $susy$ in Trouble?} So far there are no experimental signals from the $LHC$ favoring minimal models of supersymmetry, and both experimentalists and theorists are asking whether $susy$ is in trouble\cite{CMS}. By tracing the history of arguments for supersymmetry ($susy$) we find a compelling new motivation for $susy$. It is possibly the first motivation that hard-boiled experimentalists would find credible. We predict that $susy$ will soon be confirmed as a useful element of particle physics. 

Supersymmetry was invented as a way to do something beautiful with Fermions while extending the Lorentz group\cite{ramond}. Seeking beauty in mathematical creation is a long tradition in theoretical physics and also our first clue. The reality of physics is often un-beautiful. The stunning ugliness of the Standard Model happens to exist. There was never a viable phenomenology of unbroken $susy$, which is another clue. The attempt to identify Standard Model fields as super-partners of each other failed early, which is another clue\cite{BaerTata}. Almost all $susy$ particles need to be heavier than Standard Model particles, meaning that most would never be directly observable, and this is a clue. (We use ``directly observable'' here in a strict sense, as asymptotic states of the $S$-matrix, for reasons soon clear.) Once they are not directly observable, the $susy$ particles can only modify the observable correlation functions of Standard Model fields. Given current assumptions, the least massive superpartner might possibly be observable, but once again, it might just as well {\it not}.

%The same can be said for most models of new physics: a $Z'$-boson, if ever seen, would predict bumps in the invariant mass distribution of lepton pairs, but this is not expected of $susy$, and that is a clue. 

%Let us recall the great accomplishments of $susy$. It is not really surprising that the band-aid of $R$-parity conservation ``predicts'' a dark matter candidate. Every theory that hopes to have a dark matter particle needs a nearly stable particle, which every theorist knows how to concoct with a freely-constructed parity. 

The well-advertised attraction of $susy$ is improving a technical problem with renormalization of Standard Model fields. Perturbative calculations of the Standard Model\cite{sterman}, and particularly quadratic divergences of Higgs fields, show that that huge extrapolations of perturbative calculations are not reliable. By remarkable and diligent technical work, it has been shown that extending the Standard Model with $susy$ partners yields acceptable behavior of running couplings: and mainly on this point, $susy$ became popular. Minor problems such as proton decay caused a general acceptance of a seemingly artificial symmetry known as $R$ parity conservation. The decision to give $susy$ particles and Standard Model particles different $R$-parity eliminated most expectations to discover $susy$ by finding sharp bumps in invariant mass distributions. The upshot is that $susy$ has grown more and more un-discoverable. The predictions have largely merged into
slope changes, mild shoulders and kinematically smeared endpoint effects that are difficult to distinguish from Standard Model backgrounds, which is another clue.

%From Djouadi: Supersymmetry (SUSY), introduced in the early seventies by Julius Wess and Bruno Zumino
%[4, 5] among others [6] mainly for aesthetical reasons, is presently widely considered as the most
%attractive extension of the SM. The main reason is that it solves, at least technically, the hierarchy
%and naturalness problems [7]. Indeed, this new symmetry prevents the Higgs boson mass from
%acquiring large radiative corrections: the quadratic divergent loop contributions of the SMparticles
%are exactly canceled by the corresponding loop contributions of their supersymmetric partners
%which differ in spin by 1
%2 . This cancellation thus stabilizes the huge hierarchy between the GUT
%and the electroweak scales and no extreme fine-tuning is required. Later on, two other main
%motivations for introducing low energy supersymmetry in particle physics were recognized: the
%satisfactory unification of the gauge couplings of the electromagnetic, weak and strong interactions
%at the GUT scale [8] and the presence of a particle that is massive, electrically neutral, weakly
%interacting, absolutely stable, which is the ideal candidate for the dark matter in the universe [9].

\paragraph{Why Trust A Quadratic Divergence to Begin With?} The motivation for $susy$ is clear: it stabilizes an illness of perturbation theory found in the Standard Model. If this idea was once new it is now old: it is particularly unimaginative to keep repeating it, as if any other new ideas had become taboo.

New ideas and interpretations should not be taboo. We ask whether {\it perturbation theory itself might be giving us false guidance.} Perturbation theory is not designed to handle problems with widely separated time or energy scales. The outputs might be like a faulty computer operating system that gives some correct answers, and some false ones, without reliable messages of system error. When the question is raised there is a``patch.'' The patch says that perturbation theory using Feynman diagrams is the only thing known to be systematic and practical\footnote{Lest a claim be made that the renormalization group controls a hierarchy, and is in principle exact, there is actually little to trust in extrapolating inexact approximations of the renormalization group over many orders of magnitude.}. Therefore it has to be our guide, take it or leave it. While that might be used cynically, we present a new idea which accepts the main motivation for $susy$, along with the practicality of Feynman diagrams, and our idea does not need a patch.

%That comes from beginning quantum mechanics, whose structure is well-understood, and which was adopted wholesale without revision, while adding infinitely more degrees of freedom to make quantum field theory

%Anyone in the current age can make Taylor series with magic powers unimaginable 500 years ago. 
\paragraph{The Future Feynman Approach:} Imagine a future civilization fully aware of Standard Model physics and how to solve it. The infrared troubles of 21st century $QCD$, it's fudges and $K$-factors, will have long been avoided in a satisfactory way\cite{sterman}. (In order to offend no-one, we'll suggest that the magnificent collective work on $QCD$ radiative corrections and simulations is the {\it main reason} to believe in $QCD$, not the clich\'e called the ``running coupling.'' It is a shame that perturbative $QCD$ happens to have numerically important integration regions that are nothing like the actual complex analytic singularities and phase relations of physical pions, kaons and protons.) Future physicists will also have resolved the technical bother of every hierarchy problem, and their running coupling constants will be better than ours. In the future civilization there will be no issue of what correlation functions are actually measured in experiments, due to sharp awareness that very few $S$-matrix elements are actually observable, given the limited number of stable particles (photons, protons, electrons...) that actually appear in the lab. Thus the future theory in its nature and function will be more ``effective'' than we can conceive. 

If we cannot guess the non-perturbative methods of a future time, we can guess that future physicists will also have handy ways for simplified approximate calculations. Recall Feynman's trick\cite{feynman} when he found problems with covariant diagram calculations with non-Abelian theories. Years before formal justification was available Feynman tinkered with the theory by adding fake particles (``ghosts'') occurring only in loops and customized to cancel out the more obvious flaws. Eventually Feynman's semi-empirical process was justified\cite{fadeev}. Supposing those future physicists are smart and practical, they will think like Feynman, and probably introduce auxiliary fields (like Feynman's ghosts) to emulate problem issues. By construction those extra fields will have no physical reality and not propagate into the final state. It is likely those simplified calculations will be recognizable to us,
because Gaussian functional integrals are so handy they will never go out of fashion, and there are few other options to fix mishaps of perturbation theory than do more of it. 

\paragraph{Guessing Future Physics} {\it A priori} we would have little idea what kind of auxiliary fields our future colleagues will use. We don't know how to couple them to the observable particles, and even the most basic features of integrating over them is intimidating. Fortunately we have another clue. Over the past few decades hundreds of models have been explored. The consensus of experts producing brilliant technical work is that some kind of $susy$ auxiliary fields are needed to produce acceptably stable perturbation theory. Re-iterating Feynman's approach, anyone could concoct Fermion diagrams to cancel Boson divergences at one-loop order. Making the same work order by order needs a symmetry relating Bosons and Fermions. Any algebra transforming the two revises the Lorentz group algebra, hence falls into the class of supersymmetry. There is even a precedent of technically needed supersymmetry in $BRST$ gauge ghosts\cite{brst}, the ultimate successor to Feynman's tinkering. We say: {\it fine}, let us accept $susy$'s merits and calculations at face value without the unjustified and untestable naive picture they need any physical reality. Remember: {\it there is not a scrap of actual evidence that the Standard Model itself is either unstable or inordinately coupled to high energy fields.} The putative instability lies only in the unphysical ultraviolet divergences of crude approximations pushed over 14 orders of magnitude.

What is wrong with a practical calculational purpose for $susy$? Everyone working in $susy$ phenomenology knows the approach cannot be ruled out. As soon as one parameter region might be extinguished experimentally, there are 150 or more parameters to go, and in many variations. In the classic sense of testing a theory by falsifying it $susy$ is rather like strings, and untestable. What's so bad about embracing this?  The known facts support our idea: {\it you cannot rule out a new method of applied mathematics by doing physics experiments.} A mere 100 years ago the trick of representing one field by two plus a calculable Gaussian integral would have been considered futuristically sophisticated. It is no longer sophisticated, and the future physicists will be able to handle a variety of model Universes via a nearly formless, multipurpose $susy$ auxiliary field formalism with many parameters. None of these parameters will come ``from Nature.''  They will all be parameters designed to re-arrange perturbation theory and tune its flaws into representing the original theory. 

Given abundant $susy$-model calculations we think the future is not far away. It needs only a modest adjustment of discovery attitude to discover $susy$ in the near future.  

\section{Illustrative Examples}

We present some examples exploring our idea with calculations. 

\paragraph{First Example: Convergence} We notice that improving series approximations by introducing free parameters is a common technique in applied mathematics. The Gamma function makes a good illustration: \ba \Gamma( z+1) =\int_{0}^{\infty} dt \, t^{z}e^{-t}, \nn \ea For a physical analogy we can imagine a functional integral over a field $t$ with an action $e^{-t}$ to find the correlation function of $t^{z}$. Change variables to $t=zt'$, which seems harmless, but actually leads to difficulties we will discover. The result is  \ba 
\Gamma( z+1)  =z^{z}\int_{0}^{\infty} dt' \, t^{'z}e^{-zt'}  \nn \\ = z^{z}\int_{0}^{\infty} dt' \, e^{zlog(t') -zt'}.\ea  
The integrand has its maximum at $t=1$, and for large $z$ is well approximated by a Gaussian with width $1/\sqrt{z}$. The Gaussian defines a free theory for the field shifted by its $vev \, <t>=1$. The Gaussian
integral is the {\it zeroth order approximation in perturbation theory}. Powers of $(t-1)$ in the exponent are expanded  in a power series around the zeroth order term, {\it exactly as in perturbation theory: }\ba e^{zlog(t') -zt'} \sim e^{-z} e^{-(t-1)^{2}z/2  } (1+c_{3}(t-1)^{3}+c_{4}(t-1)^{4}...) \nn \ea  The Gaussian integral gives $\sqrt{2 \pi} e^{-z}/\sqrt{z}$ for the first term. Higher order terms produce the expansion known as the Stirling series \ba \Gamma( 1+z) \sim \sqrt{2 \pi} z^{z-1/2} e^{-z} (1+    {1 \over 12 z }+  {1 \over 288 z^{2} }\nn \\ + {139 \over 51840 z^{3} }+...)\label{stir} \ea Just as in field theory, this series is {\it asymptotic}, meaning it does not converge as more terms are added.  For any given $z$ only a certain maximum number of terms improves accuracy, after which adding more terms makes accuracy worse. 
 The error in truncating the series is of order the first omitted term. Every single term increases with large $z$ due to the $z^{z}$ prefactor: keeping any particular number of terms always produces an arbitrarily large error as $z \ra \infty$. The perturbative series of any given order contradicts those of any other order: this is the large-$z$ Gamma function hierarchy problem, although mathematicians do not ask Nature to solve it by changing the integral.  
 
A more subtle series approximation developed by Lanczos\cite{lanczos} and cast into our language goes as follows.  In the integral, the scale of  the field ``$t$'' is connected to renormalization group parameters, and should be extracted early. Extract the scale with a change of variables: \ba \int dt \, t^{z}e^{-t} =\mu^{z+1} \int {dt\over \mu} \, (t/\mu)^{z}e^{-\mu (t/\mu)}. \nn \ea Now $\Gamma(1+z) =A(z, \, \mu) B(z, \mu)$ where $A(z, \, \mu)= \mu^{z+1}$ and $B(z, \mu)$ is the integral. The left hand side predicts $(\mu \p \Gamma /\p\mu)=0$ because the Gamma function is ``physically observable''. The $\mu$ dependence of approximations on the right hand side creates a powerful tool. After three variable changes, and manipulation tricks characteristic of the genius of Lanczos, a series expansion emerges: \ba \Gamma(1+z) =\sqrt{2 \pi} (z+\gamma+1/2)^{z+\gamma+1/2} e^{-z-\gamma-1/2} A_{\gamma}(z), \nn \\ \label{lanc} \ea where $\gamma = 1/(1+\mu)$, and $A_{\gamma}(z)$ is a certain expansion with known coefficients\ba A_{\gamma}(z)={1\over 2}\rho_{0}+\rho_{1}{z \over z+1}+ \rho_{2}{z-1 \over (z+1)(z+2)} +...\nn \ea For completeness we list \ba \rho_{k} &=& \sum_{n}^{k} \, C_{2k}^{2n} \, F_{n}; \nn \\ F_{n} &=& {\sqrt{\pi} \over 2}(n-1)!(n+1/2+\gamma)^{-n-1/2}e^{n+\gamma+1/2}; \nn \\ cos( 2n \theta) &=&\sum_{n}^{k} \,  C_{2k}^{2n}cos^{2n}(\theta). \nn \ea Unlike Eq. \ref{stir}, the Lanczos series is {\it convergent} for all $Re(z)>-\gamma$. It is not just convergent, it is rapidly convergent. Setting $\gamma=1.5$ and keeping just two terms in the series gives \ba \Gamma(1+z) = \sqrt{2 \pi}(z+2)^{z+1/2} e^{-z-2}(0.999779+{1.084635 \over z+1}). \nn \ea This approximation has a relative error of less than $2 \times 10^{-4}$ everywhere in the right half complex plane. Keeping 7 terms with $\gamma=5$ has a relative error of less than $2 \times 10^{-10}$ over the same region. 

There are similar results for an infinite number of free parameters $\gamma$. What is even more impressive is found by taking the ``ultraviolet cutoff'' $\gamma \ra \infty$. The series simplifies and the {\it Lanczos limit formula} is \ba \Gamma(1+z) &=& \lim_{\gamma\to\infty}   \, \gamma^{z}( {1\over 2} -e^{-1/\gamma}{z \over z+1}-e^{-4/\gamma}{z-1 \over ( z+1)(z+2)}+ ...), \nn \\ &=&\lim_{\gamma\to\infty} \,  2 \gamma^{z} \, \sum_{k}^{\infty} \, (-1)^{k}e^{-k^{2}/\gamma} { (\stackrel{z}{k} )  \over (\stackrel{z+k}{k}  )} .\nn \ea The result is {\it exact everywhere} in the complex plane.  And so it is {\it very intelligent} to introduce non-existent variables and manipulate their non-existence to improve approximations.

\paragraph{Second Example: False Signals from Heierarchy} Consider the Hermitian eigenvalue problem \ba (H_{0}+ \lambda V) |\ket \psi>_{n}= E_{n} |\ket \psi>_{n},\nn \ea where the eigenvalues $E_{n}^{(0)}$ and eigenstates of $H_{0}$ are known.  Rayleigh-Schroedinger perturbation theory proposes a series expansion \ba E_{n} = E_{n}^{(0)}+ \lambda V_{nn}+\lambda ^{2} \sum_{m} \, V_{nm}{1\over  E_{n}^{(0)}- E_{m}^{(0)} }V_{mn} +...\label{ppt}\ea The matrix elements $V_{nm}=<n|V|m>$ are evaluated in the zeroth order states. Cases where $\lambda \gtrsim 1$ are called ``strong coupling'' and the series is generally recognized as worthless. We expect readers recognize that field-theoretic perturbation theory has just the same character, buried many degrees of freedom. It is not so well recognized that the interaction of a high-energy system with small coupling constant $\lambda<<1$ is also a ``strong coupling'' problem, if the limit of ``high energy'' is pushed far enough.

To explore this, let the eigenvalues of $H_{0}$ be of order ``one'' in our units, and acceptably close to the exact  ``small'' eigenvalues of $H$.  Let the exact $H$ have some ``large'' eigenvalues of a scale $E_{gut}>>1$. Let the small and large scale systems communicate by matrix elements $V$. Consider the second-order expression of Eq. \ref{ppt}. Unless a symmetry prevents any coupling at all, some couplings $V_{nm}$ generically range up to $E_{gut}$, because the scale is so large. In other words, threats to a hierarchy are generic. (See the analytic explanation of this below). The rough scaling of the second order correction in Eq. \ref{ppt} is \ba \Delta E_{(2)} \sim \sum_{nm} \lambda^{2}{E_{gut}^{2} \over E_{n}^{(0)}- E_{GUT}  } \sim \lambda^{2} E_{gut} \nn \ea Unless something special prevents it, high energy sectors {\it treated in perturbation theory} tend to push high energy into lower energy systems. But is that phenomenon reliable?

What is reliable depends on the order of limits. Perturbation theory may be well-motivated for $\lambda \ra 0$ at $E_{gut}$ fixed. Yet the series expansion fails for $\lambda=fixed<<1 $ when $E_{gut} \ra \infty.$ The energy hierarchy problem has limit-interchange features which make it unsafe to ever take $E_{gut} \ra \infty.$ 

\begin{figure}[htbn]
\begin{center}
\includegraphics[width=3.5in]{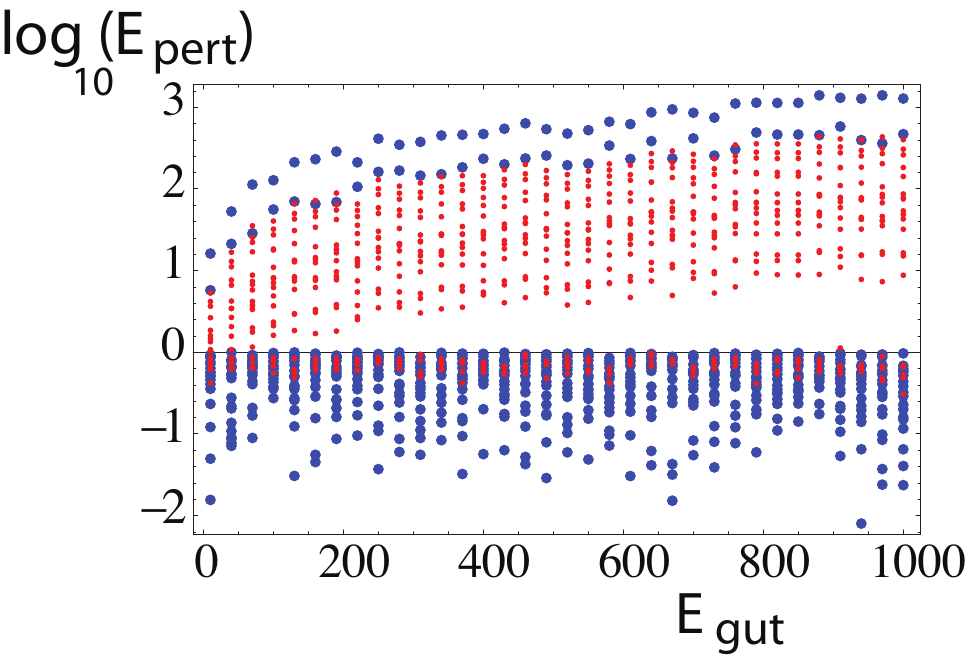}
\caption{ \small Perturbative eigenvalues of a $20 \times 20$ matrix  $H_{tot}(E_{gut)}$. The exact eigenvalues (large points, blue online) consist of two large numbers of order $E_{gut}$ and 18 small numbers of order 1. The perturbative eigenvalues (log-10 scale, small points, red online) are spread over the whole range $0<E<E_{gut}$. Scatter is due to randomizing the subspace containing the large eigenvectors at each step.  }
\label{fig:LogPert20-5-2.eps}
\end{center}
\end{figure}

Some calculations illustrate this in more detail. Consider $N=20$, and let the exact mass spectrum have 18 small eigenvalues of order 1 and 2 large eigenvalues of order $E_{gut}$. 
%The Standard Model is represented by $m_{0}$ occupying a 5-dimensional subspace, where $H_{0}$ is diagonalized.
  Treat $E_{gut}$ as a free parameter, and compute the first and second order perturbative predictions for its eigenvalues. Fig.\ref{fig:LogPert20-5-2.eps} shows the results. (The vertical scale in the figure is logarithmic). The perturbative energies as a function of $E_{gut}$ are spread over the whole range $0<E<E_{gut}$ for almost all states. This is a ``hierarchy problem.''  It is also a very poor representation of the exact eigenvalues shown in the figure, which always consist of 2 large and 18 small numbers.  

\begin{figure}[htbn]
\begin{center}
\includegraphics[width=3.5in]{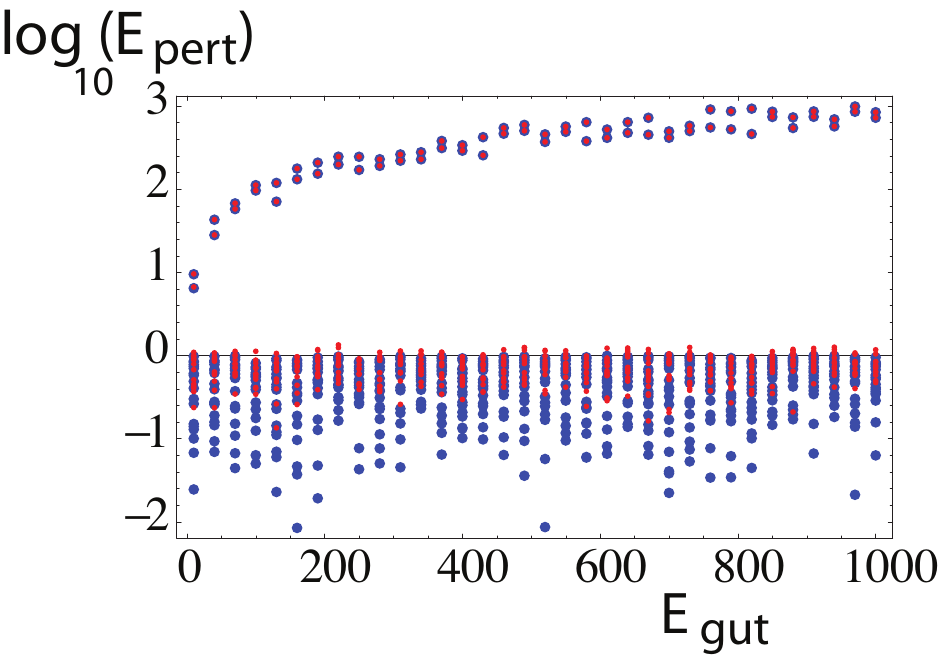}
\caption{ \small Same as Figure \ref{fig:LogPert20-5-2.eps}, except perturbing about an exact solution to the ``high energy'' sector containing the large eigenvalues. The large eigenvalues are good, while small eigenvalues develop errors of relative order 1. }
\label{fig:LogPertQQ20-5-2.eps}
\end{center}
\end{figure}

In a complementary study the complementary high-energy block  is diagonalized. We do this because diagonalizing the largest matrix elements might be a better procedure than neglecting them.  Fig. \ref{fig:LogPertQQ20-5-2.eps} shows that the perturbative calculation is relatively good. The number of large eigenvalues is trivially correct, and the numerical values are good. The flow of low energy to high energy in perturbation theory causes no problem for high energy. Perhaps future physicists will do something intelligent with the high energy sector that would allow easy hierarchies moving down. Not surprisingly, the small eigenvalues develop errors of relative order 1: the low energy sector incorrectly gives up an arbitrary fraction of its energy, but it does not run away. Both cases suggest that the worries of instability (excessive coupling to hidden freedoms) when there is an energy hierarchy are due to bad approximations.

It is possible to complain that conditions of self-consistent perturbation theory were not considered in taking $E_{gut}>>1$.  That's not a problem invented by our analysis. It is the problem and the fact of blindly computing divergent quantities (very large numbers) in perturbation theory with $E_{gut} \ra \infty$, while staying focused on particular reference parameters called coupling constants. The perturbative limit of $couplings \ra 0$ with $E_{gut}=fixed$ is seldom the same as $couplings=fixed$ and $E_{gut} \ra \infty$.

\paragraph{Third Example: Hierarchal Self-Consistency} We contrast the perturbation theory with a interesting solvable model. Consider a set of $N$ quadratically coupled fields $\phi_{i}$ with action \ba S =\int d^{4}x \, \sum_{i} \, {1\over 2}\p_{\mu } \phi_{i}  \p^{\mu } \phi_{i}-{1\over 2} m_{ij}^{2} \phi_{i}\phi_{j} ,\label{lag} \\ =\int d^{4}x  \, {1\over 2}\p_{\mu }\phi \cdot \p_{\mu } \phi-{1\over 2}  \phi \cdot  m \cdot   \phi.\nn \ea Just as in the Standard Model, a selected low-energy sector of this theory defines a Lagrangian $L_{0}$, which is solved when decoupled. The low-energy sector will be defined by using only a restricted set of low mass fields $\phi
^{(0)}=P \phi$, where $P=P^{2} $ is a projector of a certain rank $dim(P)$. The remaining fields will be called ``gut-scale'' 
and defined by $\phi^{(gut)} =Q\phi$, where $Q^{2}=Q$, $QP=PQ=0$, $P+Q=1$. The full Lagrangian is \ba L&=& L_{0}+L_{int}+L_{gut}; \nn \\ L_{0}&=&  {1\over 2}\p_{\mu }\phi \cdot P \cdot   \p^{\mu } \phi -{1\over 2} \phi \cdot P \cdot m \cdot P \phi; \nn \\ L_{int}&=& 
 -{1\over 2} \phi \cdot ( Q \cdot m \cdot P  -P\cdot  m\cdot Q)\cdot  \phi; \nn \\ L_{gut}&=& 
{1\over 2}\p_{\mu }\phi \cdot Q \cdot   \p^{\mu } \phi -{1\over 2} \phi \cdot Q  \cdot m \cdot Q  \phi. \nn \ea 

We specify that the exact system has a hierarchy where the exact spectrum\footnote{The simplicity of our model might be misunderstood as an ``exceptional theory.'' The most general quantum field theory is just as simple: quantum mechanics always boils down to diagonalizing a Hamiltonian operator on the underlying Hilbert space.} of masses falls into two groups. As before the``standard model'' or ``observable'' sector will have small eigenvalues of order $E_{small} \lesssim 1$ in low-scale units of 100 GeV, say. The ``gut'' sector will large eigenvalues of order $E_{gut}>>1$.  The ``number of large (small) dimensions'' will mean the number of eigenvectors associated with large or small eigenvalues. However we do not know the exact eigenvalues nor the number of small scale and gut-scale fields {\it a priori}, because we have not solved the interaction. 

How shall we choose the approximate low energy projections defining $L_{0}, \, L_{int}$, and $L_{gut}$? Since we are interested in generic features we will consider random systems. A random matrix of a given system is found by making a coordinate transformation with a random unitary operator $U_{random}$: \ba  m_{random}  =U_{random}\cdot m   \cdot U_{random}^{\dagger}. \nn \ea A random mass matrix of a given system is defined by \ba m_{PP}= P \cdot m_{random} \cdot  P. \nn \ea  The matrix $U_{random}$ will be distributed by the invariant Haar measure using standard numerical codes.

\begin{figure}[htbn]
\begin{center}
\includegraphics[width=3.5in]{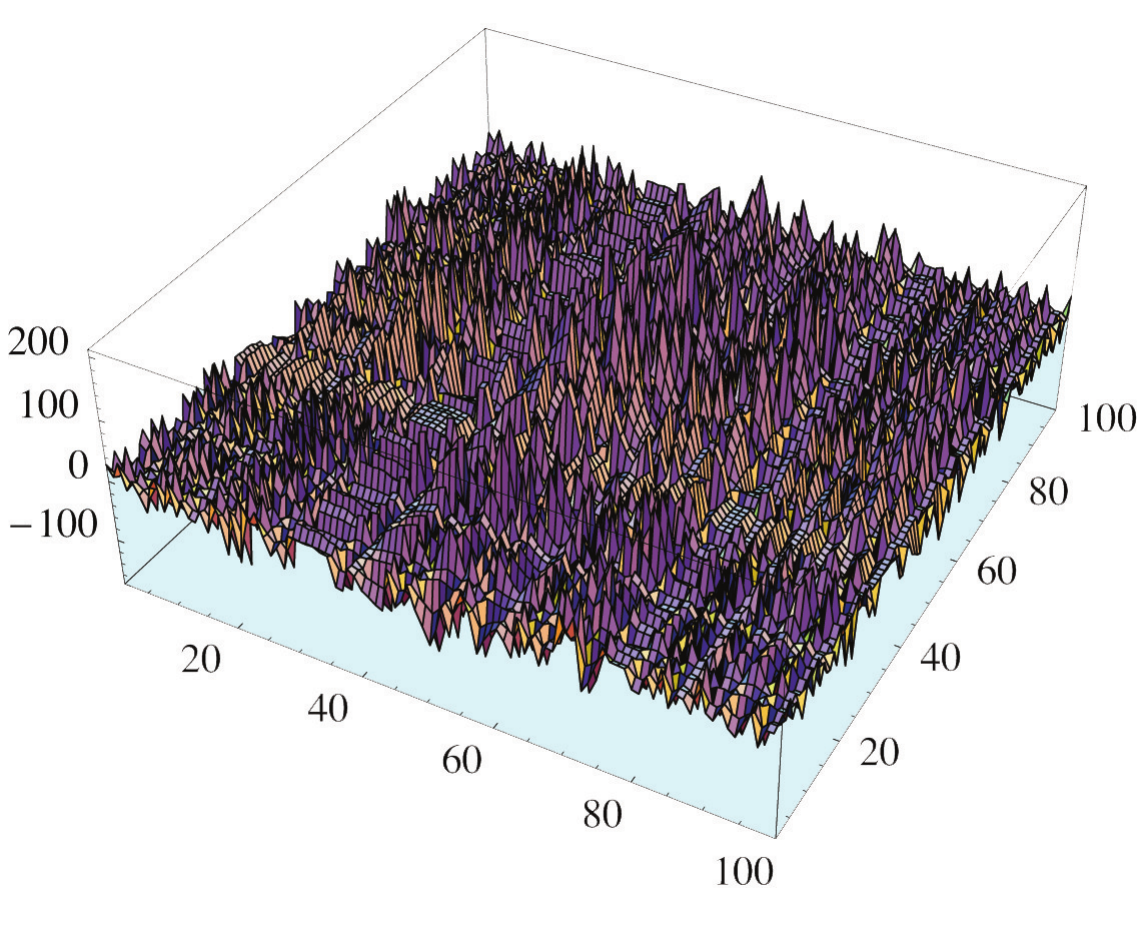}
\caption{ \small Three-dimensional plot of the random matrix elements $m_{ij}$ of a $100 \times 100$ matrix. One eigenvalue 
$E_{gut}=10^{3}$; the other 99 eigenvalues are random numbers between 0 and 
1.  }
\label{fig:100Surface.eps}
\end{center}
\end{figure}

We now come to the root of the {\it perceived} hierarchy problem. Random diagonal and off-diagonal matrix elements of systems with large hierarchies are large.  Fig.\ref{fig:100Surface.eps} shows a three-dimensional plot of the matrix elements $m_{ij}$ of a random $100 \times 100$ mass matrix. One eigenvalue $\lambda_{gut}=E_{gut}=10^{3}$; the other 99 eigenvalues are random numbers between 0 and 1. A single large eigenvalue is sufficient to cause the matrix elements to be distributed over the whole energy range with a width $\sigma_{m} \sim E_{gut}/N$. The off-diagonal distributions are centered on zero, while the diagonal elements are centered on $ E_{gut}/N$, which can be anticipated by considering $tr(m) \sim E_{gut}$. The arbitrariness of choosing an appropriate ``low energy subsystem'' creates concerns of {\it naturalness} and {\it fine tuning}. When the hierarchy is large it appears that any sub-matrix with self-consistent small eigenvalues must require a great deal of {\it fine tuning.} The lore of fine-tuning suggests we must choose the projector $P$ very carefully to get small eigenvalues. Once chosen, the lore maintains the subsystem is either kept from mixing with repeated fine-tuning, or protected by profound symmetries. 

Those worries are false, and they come from false calculations of perturbation theory. A example will illustrate this. Consider a system with the very large dimension 7 (to accommodate typesetting). Let the exact mass eigenvalues be six random numbers 
between 0 and 1 (observable sector), and one large number of order $10^{3}$ (gut
sector.) The actual eigenvalues selected were (0.586, 0.231, 0.219, 0.161, 0.119, 0.109, 
999 ).   Generate a random real matrix having these eigenvalues: \begin{small} \ba m_{tot}=
\left(
\begin{array}{lllllll}
 {80.7} & -{24.48} & {165.6} & {67.91} & \vline
   {115.5} & {165.7} & -{21.58} \\
 -{24.48} & {7.667} & -{50.22} & -{20.6} &
   -{35.01} & -{50.13} & {6.495} \\
 {165.6} & -{50.22} & {340.8} & {139.8} &
   {237.6} & {341.} & -{44.55} \\
 {67.91} & -{20.6} & {139.8} & {57.52} &
   {97.5} & {140.} & -{18.33} \\  
 {115.5} & -{35.01} & {237.6} & {97.5} &
   {165.9} & {237.8} & -{31.08} \\
 {165.7} & -{50.13} & {341.} & {140.} &
   {237.8} & {341.7} & -{44.77} \\
 -{21.58} & {6.495} & -{44.55} & -{18.33} &
   -{31.08} & -{44.77} & {6.148}
\end{array}
\right) \nn  \ea \end{small}  Choose the low energy sector to be 4 dimensional, because we want 3 generations of low 
energy, plus one window into TeV-scale new physics. Yet many matrix elements are large.  How will we find an appropriate low energy subspace without searching through every possible subspace? 

The answer lies in {\it common experimental self-consistency}. Ask an experimentalist to arbitrarily select the upper-left $4 \times 4$ block as 
a trial low-energy subspace. The eigenvalues are 0.300463, 0.183553, 0.118482, 
486.088. Note the hierarchy: 
somehow the arbitrarily selected sub-matrix has 3 small eigenvalues, without any
 ``fine tuning''.  

The result appears to be a fluke. Choose the lower-right $4 \times 4$ block.  It has 
eigenvalues (0.366832, 0.194968, 0.142922, 570.548). Note the hierarchy.

Repeat this experiment 10,000 times, with $E_{gut}=999$ fixed, while using random small eigenvalues distributed over the interval 0-1.  (If not specified otherwise, ``random'' numbers come from a flat distribution.)  Save the eigenvalues of an arbitrarily selected $4 \times 4$ block each time. {\it Every single run shows a hierarchy with 3 
small eigenvalues and one large one.} In more detail: the mean eigenvalues are (0.3, 0.5, 0.7, 570), with standard deviations (0.15, 0.15, 0.15, 205).  
The $p-value$ probability to find any large eigenvalue {\it less than 100} is about $6 \times 10^{-3}$.  There were no events with any of the three small eigenvalues greater than 1.  

\begin{figure}[htbn]
\begin{center}
\includegraphics[width=3.5in,height=4in]{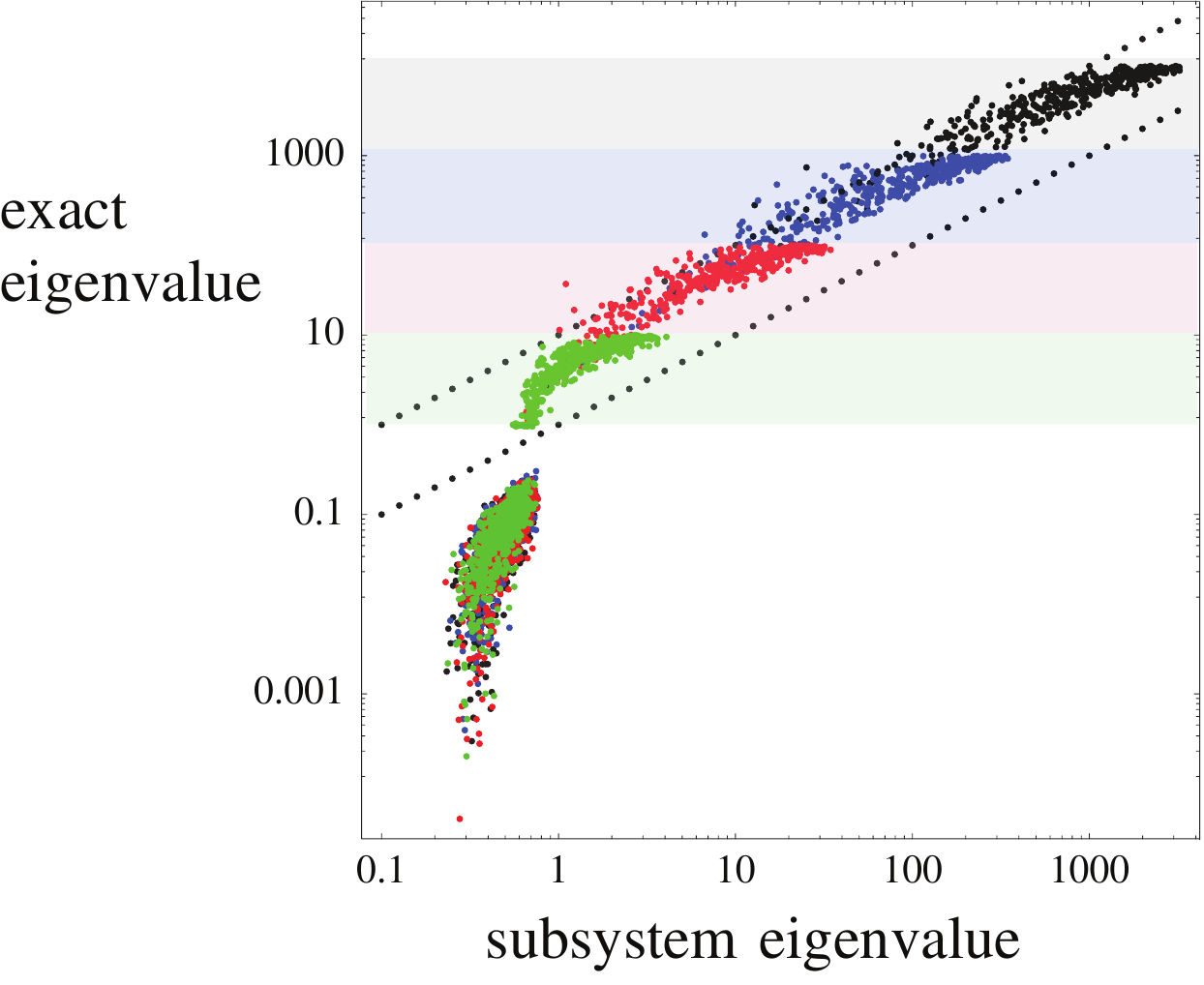}
\caption{ \small Self-consistent hierarchies by correlation of random subsystem eigenvalues with exact eigenvalues. Exact large eigenvalues are 4 random numbers times $E_{gut}$. Subspace eigenvalues are found to be 4 large numbers in almost every random case. 
Different studies use $E_{gut}=$ 10, 100, 1000, and 10,000, producing the correlations seen (color online). Color bands indicate upper range of each study. Study uses $dim(m_{tot})=50,\, dim(P)=10$.  }
\label{fig:HierarScaling.eps}
\end{center}
\end{figure}

There is nothing special about $4 \times 4$ projections of $7 \times 7$ matrices and the same phenomenon is found in thousands of examples. The mathematical literature on random matrices is dominated by finding invariants (eigenvalues), {\it given} random matrix elements. For physics we need the inverse problem of characterizing matrix elements that represent couplings, {\it given} invariants. The {\it theory of eigenvalue hierarchy} has not been explored this way before, and it is fascinating\cite{john}. In brief, hundreds of experiments led to two remarkable regularities: \begin{itemize} \item We find an empirical {\it conservation law of large eigenvalues,} which is that (1) the number of large eigenvalues is conserved on almost any subspace, and (2) their magnitudes scale like $E_{gut}$. {\it The exact large eigenvalues can even be predicted in a statistical sense from the eigenvalues on a subspace.} Fig. \ref{fig:HierarScaling.eps} shows the correlation of subsystem eigenvalues with exact eigenvalues. \item There is a more detailed ``scaling law'' of spectral similarity. Let $\xi =dim(P)/dim(m)$ be the ratio of dimensions reducing to the subsystem. Then ordering the eigenvalues by size, and normalizing the sum to one, the exact spectrum $\lambda(k)$ for state $k$``tries to be'' reproduced in a scale-similar way on the small system with eigenvalues $\lambda_{\xi}(k) $: \ba \lambda_{\xi}(k) =\xi \lambda ( \xi k). \label{sim} \ea  Eq. \ref{sim} is an approximate fact. It is violated for small numbers of eigenvalues by the law of conservation of large eigenvalues.  \end{itemize} 

Our proposal to explain these observations is geometric. The exact eigenvalues of $m$ are extrema of $<\psi|m|\psi>$ over normalized $<\psi|\psi>$. The eigenvalues of $P \cdot m\cdot P$ are the extrema of $<\psi|P \cdot m\cdot P|\psi>$.  The locus of these expectations are generalized ellipsoids in high-dimensional spaces.  An ellipsoid projected onto a subspace is an ellipsoid partaking of the dimensions found in the subspace. When there is a great hierarchy of eigenvalues on a big space, it produces much the same hierarchy on almost every subspace, or ``the shadow of a needle is a needle.'' 

We have provided non-perturbative evidence that fine-tuning and naturalness concerns over energy hierarchies do not exist. We conclude this section by remembering that experimentalists easily diagonalize subsystems by measuring the spectra of masses and energies in the lab. The experimental low energy sector defines itself self-consistently, and without needling any dynamical conspiracy. Yet reproducing that outcome using perturbation theory absolutely needs significant theoretical re-arrangement, with integration over auxiliary fields being the most attractive method.

\section{Predictions for Today's Susy Searches} Can we use these observations to do something productive in present day? We do not know where the important perturbative errors actually occur. However Nature will do the calculation exactly, so we should take our guidance from experiment. Many groups are currently fitting $susy$ model parameters\cite{enrico}, without finding any evidence they are on the right track. I observe these fits are based on the needless assumption $susy$ is new physics. When we use $susy$ as an effective theory of the Standard Model expressed with auxiliary fields, each set of $susy$ parameters needs to be self-consistently adjusted to the particular theory defects it is designed to ameliorate. For example, the anomalous magnetic moment $g-2$ of the muon\cite{Brown:2001mga} provides a very important constraint on ordinary parameter fitting. Other strong constraints included in current studies include branching ratios\cite{heavyflavor} such as $\bar B \ra X_{s}\gamma$, $B_{s} \ra \mu^{+}\mu^{-}$, and $B_{u} \ra \tau \nu$. Yet when fitting $LHC$ data, where the uncertain theoretical issues lie largely in perturbative $QCD$, Monte Carlo simulations, unknown correlations of jet substructure, etc, it is contradictory to bring in $B \ra s \gamma$, the muon's $g-2$, etc. In fact, there is more than one reason why the discrepancies of $g-2$ may well lie in Standard Model physics not needing any $susy$ corrections at all\cite{rainer}. In recognition that $g-2$ is suspicious\cite{gm2review}, some groups are already dropping it selectively to check its effects.\cite{Fowlie:2012im}.  Yet constraints from dark matter direct detection and relic abundance pose further barriers that may be entirely specious, given the full range of unknowns. With the new interpretation, {\it  I propose dropping from studies of $LHC$ data all irrelevant constraints}, and exploring whether $susy$ auxiliary fields produce much better fits.  

The proposal is scientifically conservative: it is more conservative, and modest, than assuming $susy$ must exist so that Nature can take care of theoretical approximation schemes. The point of developing models is to fit data and test theories. The theory being tested is the Standard Model. The Standard Model fits so much data so well it is difficult to improve by adding parameters. We have no intention of selecting data in order to fit a model. We suggest taking an existing, standardized $LHC$ collider physics analysis and finding if a few-parameter $susy$ effective action will improve the fit {\it at a statistically significant level}. This idea is new, since there is no prior evidence of statistically significant improvement, and it is testable when the statistical penalties of adding parameters are taken into account. For reference, one expects on statistical grounds to improve a $\chi^{2}$ statistic by about one unit ($1\sigma$)  by adding one parameter. If a statistic improves by an additional $3\sigma$ above expectations, then enthusiasts can get excited, while large experimental collaborations often cite $5\sigma$ improvement as their objective. If (say) adjusting two parameters known as the standard $(m_{0},  \, m_{1/2})$ set gives enough units of improvement to pass a pre-selected threshold of statistical significance, my prediction will find support. 

What good is the information from an improved fit? It is first of all indefensible not to make the effort to find one. Once an improved fit is found its details will give a theoretical microscope into the possible causes. The differences between interaction rates computed in a theory is always much more dramatic before applying hadronization and acceptance factors. Then comparing simpler, pre-hadronized theory to theory will help identify what features of the Standard Model calculation need help from parameters.    

Suppose $susy$ really does exist. If it cannot be falsified, can our suggestion help truthify it?  Absolutely! It is a first principle of scientific sleuthing to deal with simpler problems before taking on over-complicated ones. If a multi-sigma minimum is found in the $(m_{0}, \, m_{1/2})$ plot when omitting some of the standard but superfluous assumptions, it may well be a clue where to focus on discovery. It may also give clues where to go back and re-assess prejudices about supposedly established backgrounds and facts - such as the reliability of $g-2$ computed in perturbation theory, which we claim is not solid. Suppose new physics exists which is not $susy$. Same procedure: we claim it is generally useful to use a flexible method of applied mathematics to first establish a statistically significant fit. Once a ``signal'' exists, the details are sorted out by identifying the theory elements that distinguish it.\\

{{\it Acknowledgements:}} Research supported in part under DOE Grant Number DE-FG02-04ER14308.
We thank KC Kong, Doug McKay and Pierre Ramond for helpful suggestions. 

%%running alpha, self confirming observables that are consistent;  role of susy
%
%It is common for $susy$ advocates to point to unifying coupling constants extrapolated over many orders of magnitude. When one takes coupling constants seriously, it is possibly naive to compute their running in perturbation theory.  The running coupling computed in perturbation theory involves integration over unrealistic thresholds and singularities, and not just at new thresholds, but everywhere.  Since perturbation theory produces asymptotic series, one also has no way of knowing how many loops will give the best results. 

%http://www.scientificamerican.com/podcast/episode.cfm?id=000427A2-CC3B-1458-8C3B83414B7F0000

\end{document}